\documentclass[%
reprint,
amsmath,amssymb,
aps,
]{revtex4-2}
\usepackage{graphicx}
\usepackage{dcolumn}
\usepackage{bm}
\usepackage{hyperref}
\usepackage{xcolor}

\newcommand{\dd}{\mathrm{d}}
 
\begin{document}
	\title{Heat transfer across a vacuum gap induced by piezoelectrically mediated acoustic phonon tunneling}
	\author{Zhuoran Geng}
	\email{zhgeng@jyu.fi}
	\author{Ilari J. Maasilta}%
	\email{maasilta@jyu.fi}
	\affiliation{%
		Nanoscience Center, Department of Physics, University of Jyv{\"a}skyl{\"a}, FI-40014 Jyv{\"a}skyl{\"a}, Finland
	}%
	\date{\today}
	
	\begin{abstract}
		In contradictin to the common concept that acoustic phonons can only travel inside a material medium, they can in fact "tunnel" across a vacuum gap with the help of piezoelectricity, transmitting a significantly stronger heat flux than that of blackbody radiation. Here, we present a theoretical formulation for the heat flux of such piezoelectrically mediated heat transfer, applicable to any anisotropic piezoelectric crystals with an arbitrary orientation. A few numerical results are demonstrated and compared to heat transfer driven by other close-range mechanisms, including near-field radiative heat transfer and other acoustic phonon tunneling mechanisms. 
		We find that piezoelectrically mediated heat transfer has a significant effect when the vacuum gap size is smaller than the phonon characteristic thermal wavelength, and its heat flux can dominate heat transfer between piezoelectric solids over all other known heat transfer mechanisms at temperatures below 50 K. 
	\end{abstract}
	
	\maketitle
It is known that heat can be transfered between macroscopic bodies via three different channels, namely, conduction, convection and radiation. Out of those, radiation is the only possible channel between two materials separated by a vacuum gap, and the corresponding heat flux is well understood based on Planck's law of radiation. However, when the separation between two material bodies decreases, heat flux exceeding that of Planck's law by several orders of magnitude has been observed in experiments\cite{DOMOTOGA1970,Song2015,Lucchesi2021}. Among the mechanisms that can lead to such super-Planckian radiation, near field radiative heat transfer (NFRHT) mediated by photon tunneling\cite{Polder1971,Pendry1999,Volokitin2001,Joulain2005} is the most studied one. Analogous to quantum mechanical tunneling, this photon tunneling becomes relevant when the vacuum gap distance is below the photon thermal wavelength, which is about $\sim10\,\mu$m at room temperature. 

With advances in nanotechnology, vacuum gaps with sizes in the nanometer-to sub-nanometer range can be achieved in  experiments\cite{Kim2015,Kloppstech2017,Cui2017,Jarzembski2022}. This has stimulated active research in recent years also on the heat transfer driven by the tunneling of acoustic phonons\cite{Prunnila2010,Sellan2012,Persson2011,Chiloyan2015,Budaev2011,Ezzahri2014,Sasihithlu2017,Pendry2016,Volokitin2019,Volokitin2020,Fong2019,Biehs2020,Jarzembski2022} in addition to photons, as their thermal wavelengths are at that length scale. 
However, the concept of acoustic phonon tunneling is far from obvious, as a phonon, being a vibration of the atomic lattice, requires the presence of a medium to propagate. In the past decade, a few mechanisms that can mediate acoustic phonon tunneling have been suggested, the van der Waals force\cite{Budaev2011,Ezzahri2014,Sasihithlu2017,Pendry2016} the electrostatic force\cite{Pendry2016,Volokitin2019,Volokitin2020}, or the non-local contribution of acoustic phonons to NFRHT in polar crystals \cite{Viloria2023}. Nevertheless, these studies show that the heat flux due to these mechanisms decay rapidly with the gap width ($d^{-7}$ and $d^{-9}$ for van der Waals and electrostatic mechanisms,  respectively), and hence provide non-trivial contributions to the heat transfer only when a vacuum gap $d<1$ nm at room temperature\cite{Pendry2016}. 

There exists yet another, much less studied acoustic phonon tunneling mechanism, which utilizes piezoelectricity. A thermally excited acoustic phonon impinging on a free surface of a piezoelectric solid can create a decaying, evanescent electric field leaking into the vacuum. Such field couples to the lattice deformations of a second piezoelectric solid placed within the phonon wavelength, leading to a transmission of heat across the vacuum. Such piezoelectrically mediated heat transfer (PEMHT) was previously suggested and studied in Ref. \cite{Prunnila2010}, however the theoretical estimation in that study was based on a highly simplified model, in which the isotropic material parameters were assumed and only two phonon modes were considered. Moreover, their results are not in agreement with those derived from the more general acoustic wave tunneling formalism for piezoelectric materials developed in Refs.\cite{Geng2022_1,Geng2022_2}, and hence the PEMHT phenomenon needs to be re-examined. 

In this work, we present a general formulation for the piezoelectrically mediated heat transfer via acoustic phonon tunneling, which can be applied to arbitrarily anisotropic and oriented piezoelectric crystals. Numerical examples are investigated and compared to other close-range heat transfer mechanisms, including non-piezoelectric tunneling of acoustic phonons and near-field radiative transfer (photon tunneling). We find that PEMHT can dominate heat transfer at temperatures below 50 K. At the end, numerical examples of PEMHT using different piezoelectric materials with varying crystal orientations are discussed.  

\begin{figure}[ht]
	\centering
	\includegraphics[width=0.6\linewidth]{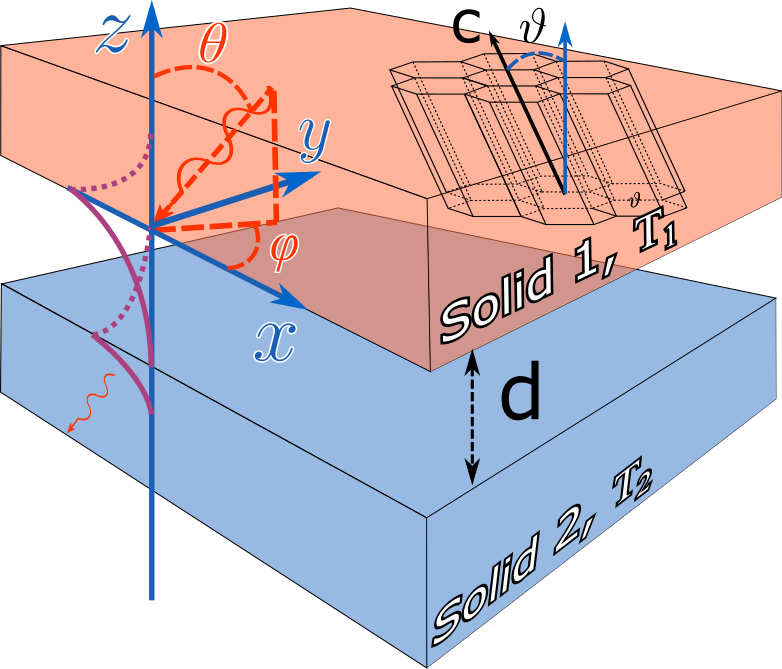}
	\caption{Two piezoelectric solids 1, 2, at temperatures $T_1$ and $T_2$, are separated by a vacuum gap of width $d$. 
    An incoming acoustic phonon (red) with an incident zenith angle $\theta$ and azimuth angle $\varphi$ tunnels across the vacuum from solid 1 and transfers heat to the adjacent solid 2. 
    In this study, we demonstrate results for hexagonal crystals, (ZnO,AlN), whose orientation can be described by a rotation angle $\vartheta$ between the crystal $c$-axis and the laboratory $z$-axis.}
	\label{fig:coordinates}
\end{figure}

We consider two piezoelectric, semi-infinite solids which are placed parallel to each other and separated by a vacuum gap of width $d$, as shown in Fig.\ref{fig:coordinates}. Both solids can be rotated to arbitrary orientations described by a set of Euler angles denoting the relation between the intrinsic crystal coordinates and the laboratory coordinates $xyz$ (see \cite{supplement} for more details on crystal orientation). In our model, these solids are assumed to be continuous and anisotropic, satisfying linear elastic constitutive equations\cite{Auld1973,Geng2022_1}.   

We assume that the thermal acoustic phonons in solid 1 impinging on the gap  consist of  three 
bulk acoustic wave modes $\alpha=1,2,3$\cite{Geng2022_1}, where each wave mode can be described by a time-harmonic displacement field
\begin{equation}
	\pmb{u}_\alpha=b_\alpha\pmb{A}_\alpha\exp(i\omega_\alpha t-i\pmb{k}\cdot\pmb{r}) ,
	\label{eq:wavefunction}
\end{equation}
where $\pmb{k}$ is the wave vector, $\pmb{r}=(x,y,z)$ is the position vector in the laboratory coordinates, $\omega_\alpha$ is the angular frequency of the phonon mode, $\pmb{A}_\alpha$ is the Stroh-normalized\cite{Geng2022_1} polarization vector and $b_\alpha$ is the dimensionless amplitude of the mode\cite{Geng2022_1}. 

In PEMHT, the emitted heat flux $J^{\textrm{PE}}$ from the solid $\gamma=1,2$ across the gap can then be expressed as
\begin{equation}
\begin{aligned}
     J^{\textrm{PE}}_\gamma(T_\gamma,d) = &\sum_\alpha\int \frac{\dd^3k}{(2\pi)^3}
    \hbar\omega_\alpha(\pmb{k})n(\omega_\alpha,T_\gamma) \\
    &\times\bigg(\pmb{\hat{n}}_\gamma\cdot\frac{\partial\omega_\alpha}{\partial\pmb{k}}\bigg)
    \Theta\bigg(\pmb{\hat{n}}_\gamma\cdot\frac{\partial\omega_\alpha}{\partial\pmb{k}}\bigg)
    \mathcal{T}_\alpha(\theta,\varphi,k,d),
\end{aligned}
\label{eq:PE_flux_general}
\end{equation}
where 
$n(\omega_\alpha,T_\gamma)=[\exp(\hbar\omega_\alpha/k_BT_\gamma)-1]^{-1}$ is the Bose-Einstein distribution describing the thermal occupation of the phonon mode of energy $\hbar\omega_\alpha(\pmb{k})$, $\pmb{\hat{n}}_\gamma$ is the \emph{outward} unit normal of the vacuum-solid interface, and $\mathcal{T}_\alpha$ is the power transmittance of mode $\alpha$ \cite{Geng2022_2}.

The term $\pmb{\hat{n}}_\gamma\cdot\partial\omega_\alpha/\partial\pmb{k}$ in Eq.\eqref{eq:PE_flux_general} describes the group velocity of the phonon wave mode $\alpha$ in the direction of the outward normal of the vacuum-interface, and can be expressed as\cite{supplement}:
\begin{equation}
	\pmb{\hat{n}}_\gamma\cdot\frac{\partial\omega_\alpha}{\partial\pmb{k}} = 
    \frac{1}{2}\frac{\sin\theta}{\rho v_\alpha(\theta,\varphi)|\pmb{A}_\alpha(\theta,\varphi)|^2}\hat{\xi}_\gamma(\theta,\varphi),
	\label{eq:normal_vg}
\end{equation}
where $\rho$ is the density of the solid. The phase velocity term $v_\alpha=\omega_a/k$ depends on the material, crystal orientation and incident angles $(\theta,\varphi)$, and can be solved from the piezoelectrically stiffened Christoffel equation [Eq.(8.147) in Ref.\cite{Auld1973}]. Furthermore, the polarization vector $\pmb{A}_\alpha$ is obtained from the normalized eigenvector of the extended Stroh matrix [Eq.(3) in Ref.\cite{Geng2022_1}], whereas for the bulk waves, $\hat{\xi}_\gamma=\pm1$ (J/m) is obtained from the Stroh-normalization\cite{Geng2022_2}, whose sign determines the energy flow direction of the phonon along the unit vector $\pmb{\hat{n}}_\gamma$.

In addition, a Heaviside step function $\Theta(f)$, which equals to unity (zero) when $f>0$ ($f<0$), is used to correctly select the phonons whose group velocities point from the solid towards the vacuum. It is important to note that in anisotropic crystals,  the direction of phonon propagation, signified by the direction of the group velocity $\partial\omega_\alpha/\partial\pmb{k}$, is generally different from the wave front direction given by the wave vector $\pmb{k}$. It is possible for a phonon that tunnels outward from solid 1 to 2 to have an inward wave vector in the normal direction, i.e. $(\pmb{\hat{n}}_\gamma\cdot\partial\omega_\alpha/\partial\pmb{k})<0$ but $(\hat{\pmb{n}}_1\cdot\pmb{k})>0$. Therefore, to fully account for all the possible incident acoustic phonons impinging on the surface towards the vacuum, we have to integrate over the complete k-space in Eq.\eqref{eq:PE_flux_general} including the inward half-hemisphere, but choosing the outward traveling phonons with the help of the Heaviside step function $\Theta[\pmb{\hat{n}}_\gamma\cdot(\partial\omega_\alpha/\partial\pmb{k})]$.

By following the methods presented in Ref.\cite{Geng2022_1}, the total tunneled power transmittance  of an incoming wave mode $\alpha$, coupling into all possible bulk modes in the second solid, takes the form\cite{Geng2022_2}:
\begin{equation}
	\mathcal{T}_\alpha(\theta,\varphi,k,d) = \frac{2\mathrm{Re}\big[r_V^{(2)}\big]\big|t_{\alpha\rightarrow V}^{(1)}\big|^2e^{-2kd\sin\theta}}{\big|1-r_V^{(1)}r_V^{(2)}e^{-2kd\sin\theta}\big|^2},
	\label{eq:power_transmittance}
\end{equation}
where $t^{(i)}$ and $r^{(i)}$ are the single surface transmission and reflection coefficients, which describe the scattering of the acoustic wave at the surface of solid $i=1,2$ as if there is no adjacent second solid. To be more specific, we denote with $t_{\alpha\rightarrow V}^{(1)}$ the coefficient of an incoming $\alpha$ mode wave transmitted into an evanescent wave (electrical potential) in vacuum from solid 1, and with $r_{V}^{(i)}$ the coefficient of an evanescent wave coming from vacuum reflected on the surface of solid $i$. For given material parameters, crystal orientations and incident wave mode $\alpha$, these coefficients are functions of the incident angles $\theta$ and $\varphi$, independent of $k$ and $d$, and can be obtained numerically using the boundary conditions at the solid-vacuum interface, following the formalism presented in Ref.\cite{Geng2022_1}. 

The net heat flux between solid 1 at temperature $T_1$ and solid 2 at temperature $T_2$ is the difference between their corresponding emitted heat fluxes, and reads as $\Delta J^{\textrm{PE}}=J_1^{\textrm{PE}}(T_1,d)-J_2^{\textrm{PE}}(T_2,d)$. Furthermore, in the limit where  the power transmittance [Eq.\eqref{eq:power_transmittance}] is set to unity and isotropy is assumed for the group velocity, given then by $(\pmb{\hat{n}}_\gamma\cdot\partial\omega_\alpha/\partial\pmb{k})=v_\alpha\cos\theta$, the heat flux Eq.\eqref{eq:PE_flux_general} simplifies to  $J_\gamma=\sum_\alpha\pi^2k_B^4T_\gamma^4/120v_\alpha^2\hbar^3$, recovering the expression for phonon blackbody radiation for isotropic  matter\cite{Swartz1989}.

\begin{figure}[ht]
	\centering
	\includegraphics[width=0.9\linewidth]{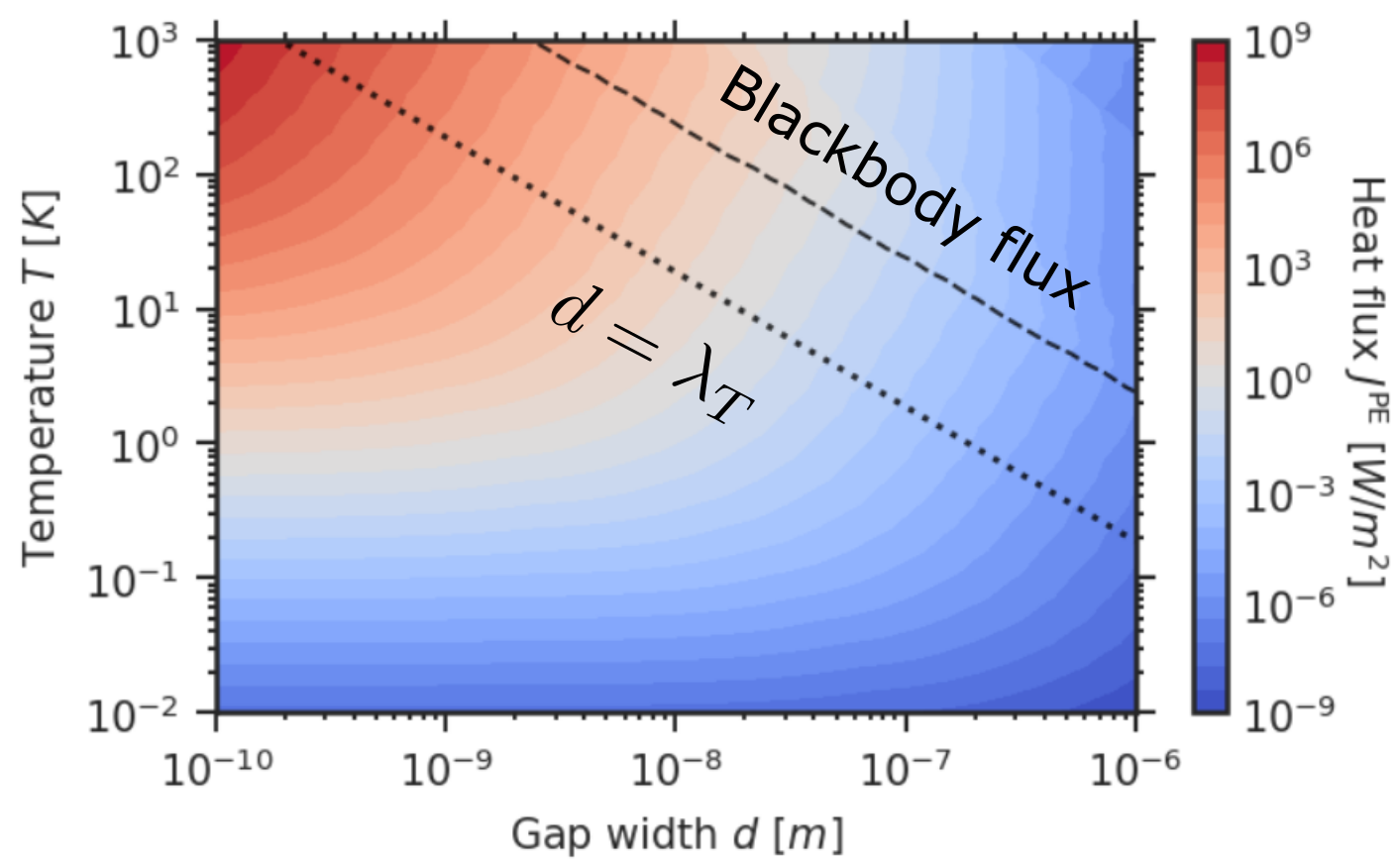}
	\caption{Contour plot of the total emitted heat flux $J^{\mathrm{PE}}$ of all phonon wave modes for ZnO as function of temperature $T$ and gap width $d$. The dotted line is $T=2\pi\hbar v/k_Bd$, denoting the condition $d=\lambda_T$ where $v=3900$ m/s. The dashed line is $J^{\mathrm{PE}}=\pi^2 k_B^4T^4/60c^2\hbar^3$, signifying the blackbody radiation limit.}
	\label{fig:fluxvstd}
\end{figure}

To demonstrate PEMHT numerically, we consider the case of two ZnO crystals with a hexagonal 6mm symmetry, with their material constants taken from Ref.\cite{Auld1973}. These solids are separated by a vacuum gap of width $d$, and are rotated identically such that their crystal $c$-axes are aligned with the laboratory $y$-axis (hence $\vartheta=\pi/2$, see Fig.\ref{fig:coordinates}). In Fig.\ref{fig:fluxvstd} we plot the emitted heat flux $J^{\mathrm{PE}}(T,d)$ between the two solids as a function of the emitter temperature $T$ and gap width $d$. By introducing the characteristic thermal wavelength $\lambda_T$ \cite{Prunnila2010,Pendry2016,Volokitin2020}, defined as $\lambda_T=2\pi v_\alpha\hbar/k_BT$, we find that the plot can be divided into two regions roughly separated by the black dotted line signifying the condition $d=\lambda_T\approx(0.2\textrm{ $\mu$m})/T$, where we used an average phonon phase velocity [$v=(\sum_\alpha v_\alpha)/3$] of $3900$ m/s for ZnO. 

Towards the lower-left region, the heat flux decreases strongly with decreasing temperature, but saturates with the gap width. This saturation comes about because only the power transmittance $\mathcal{T}_\alpha$ is a function of the gap width $d$ in Eq.\eqref{eq:PE_flux_general}, and this dependency is only expressed via the exponential term $\sim\exp(-kd)$ in Eq.\eqref{eq:power_transmittance}\cite{Geng2022_1}. As a result, for a gap width $d \ll \lambda_T\sim1/k$, 
$\mathcal{T}$ becomes constant, and the heat flux $J^{\textrm{PE}}$ is hence determined by the thermal distribution function $n(\omega,T)$, which is strongly modified by the temperature. In contrast, the exponential decay of  $\mathcal{T}$ dominates the heat flux towards the upper-right section. In the large-gap limit ($d\gg\lambda_T$), the heat flux is practically "switched-off" and becomes insensitive to the change of temperature. In addition, a black dashed line marks where the heat flux from PEMHT equals to that of the blackbody radiation at the same temperature [$J^{\mathrm{PE}}(T,d) = \pi^2 k_B^4T^4/60c^2\hbar^3$ where $c$ the speed of light]. The comparison between the dashed and dotted lines shows that PEMHT, if "switched-on" ($d\leq\lambda_T$), generally contributes at least three orders of magnitude stronger heat flux than blackbody radiation at a given temperature, making it a non-trivial source in the context of near-field heat transfer. 

\begin{figure}[ht]
	\centering
	\includegraphics[width=0.9\linewidth]{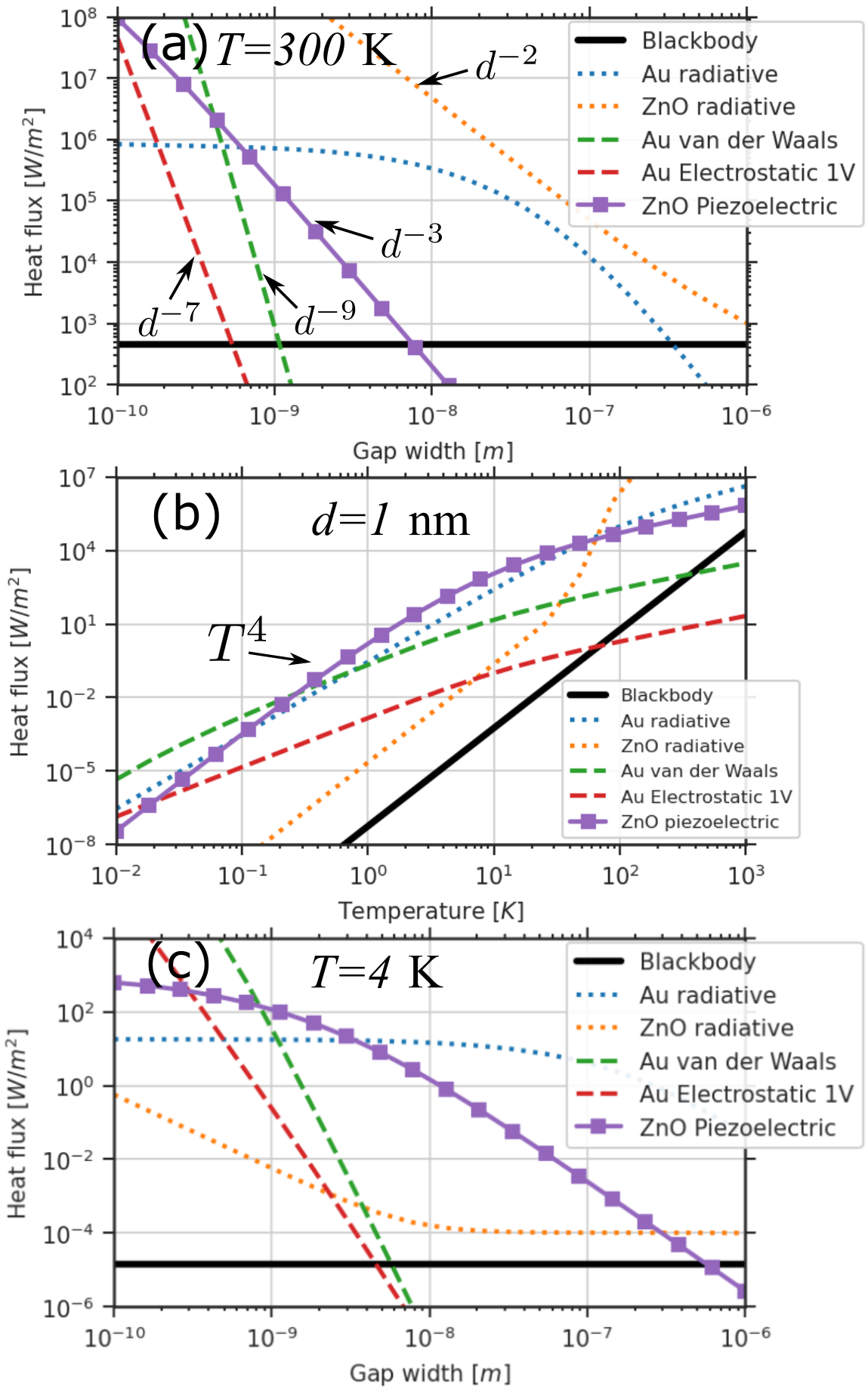}
	\caption{Comparison of the emitted heat fluxes driven by different close-range mechanisms. (a) The heat fluxes as a function of the gap width at $300$ K. Purple square marker is PEMHT of ZnO, the dotted lines are NFRHT for Au (blue) and ZnO (orange). The dashed lines are the acoustic phonon tunneling in Au mediated by van der Waals force (green) and $1$ V electrostatic potential difference (red). (b) The heat fluxes as function of temperature with a fix gap width of $1$ nm. (c) The heat fluxes as a function of gap width at $4$ K.}
	\label{fig:fluxcomparison}
\end{figure}

In addition, at room temperature, PEMHT is stronger than blackbody radiation even at a gap width close to $10$ nm. This is very different to the 
other acoustic phonon tunneling mechanisms described in literature \cite{Ezzahri2014,Pendry2016,Persson2011,Volokitin2019,Volokitin2020,Viloria2023}, for which  non-trivial heat flux occurs in the sub-nanometer length scale at room temperature. In Fig.\ref{fig:fluxcomparison}, the heat flux carried by various relevant close-range mechanisms are compared, including the near-field radiative heat transfer (NFRHT, \cite{Pendry1999,Joulain2005}) for ZnO \cite{Ashkenov2003,Ooi2011} and Au \cite{Chapuis2008}, phonon tunneling for Au mediated by van der Waals force \cite{Pendry2016} and electrostatic force (for $1$ V bias across the vacuum) \cite{Volokitin2019,Volokitin2020}, and the blackbody radiation (maximal far-field radiative heat transfer).

In panel (a) of Fig.\ref{fig:fluxcomparison}, the heat fluxes of all the above mentioned mechanisms at $300$ K are plotted as a function of the gap width $d$. It is clear that at room temperature, NFRHT for Au and ZnO, denoted by the blue and orange dotted lines, respectively, is significantly stronger than the heat transfer mediated by the acoustic phonon mechanisms. Moreover, the heat fluxes driven by the van der Waals force and the electrostatic force scale as $d^{-9}$ and $d^{-7}$, respectively \cite{Volokitin2020}, hence they only have non-trivial contributions below  $d<1$ nm at room temperature, and will fall off rapidly with the increase of the gap width. In contrast, PEMHT (square symbols) scales as $d^{-3}$, similar to that for NFRHT (photon tunneling). Consequently, PEMHT quickly dominates the other phonon tunneling mechanisms at larger-than-nanometer scales, being a relatively "long-range" phenomenon. 

One interesting observation in Fig.\ref{fig:fluxcomparison} (a) is that the NFRHT of ZnO is particularly strong. This happens\citep{Joulain2005} because the surface-phonon polaritons of ZnO can be excited at  infrared frequencies \citep{Ashkenov2003,Ooi2011}, matching the spectrum of the room temperature thermal photons, and hence enhancing the heat flux. But this also infers that one should expect a strong attenuation of the NFRHT once the excitations are stopped, \emph{i.e.} when the temperature is lowered. This is confirmed in Fig.\ref{fig:fluxcomparison}(b), in which the heat fluxes are plotted as a function of the temperature for a fixed gap width of $1$ nm. There is a clear cut-off of the NFRHT flux at about $100$ K for ZnO.

More interestingly, PEMHT becomes stronger than the fluxes from all other mechanisms, including NFRHTs, between 0.1 K to 50 K. At this temperature range, the power transmittance of the PEMHT increases exponentially as the thermal wavelengths of the acoustic phonons increase, eventually saturating at sub-Kelvin range (see Fig.\ref{fig:fluxvstd}). With the temperature lowered even further, the PEMHT flux is determined by the phonon state energy term (phonon thermal spectrum), and therefore has a $T^4$ dependence on the temperature, similar to the NFRHT of ZnO in the low temperature limit. Meanwhile, the other acoustic phonon tunneling mechanisms, scaling more slowly than the PEMHT, begin to dominate the heat transfer at the temperatures below 0.1 K.

From the above numerical analyses, we believe that PEMHT can be experimentally observed using modern nanofabrication techniques at cryogenic temperatures. As an example,in Fig.\ref{fig:fluxcomparison} (c) we demonstrate the heat fluxes from all the above mentioned mechanisms, as a function of the gap width at 4 K. 
One finds that the PEMHT of ZnO (purple squares) dominates the heat flux up to a gap width as high as $d\sim300$ nm, compared to the NFRHT of ZnO (orange dotted line) and blackbody radiation (black line). With further lowering of the temperature or reduction of the gap width, PEMHT will become more prominent and easier to observe. 

\begin{figure}[ht]
	\centering
	\includegraphics[width=0.8\linewidth]{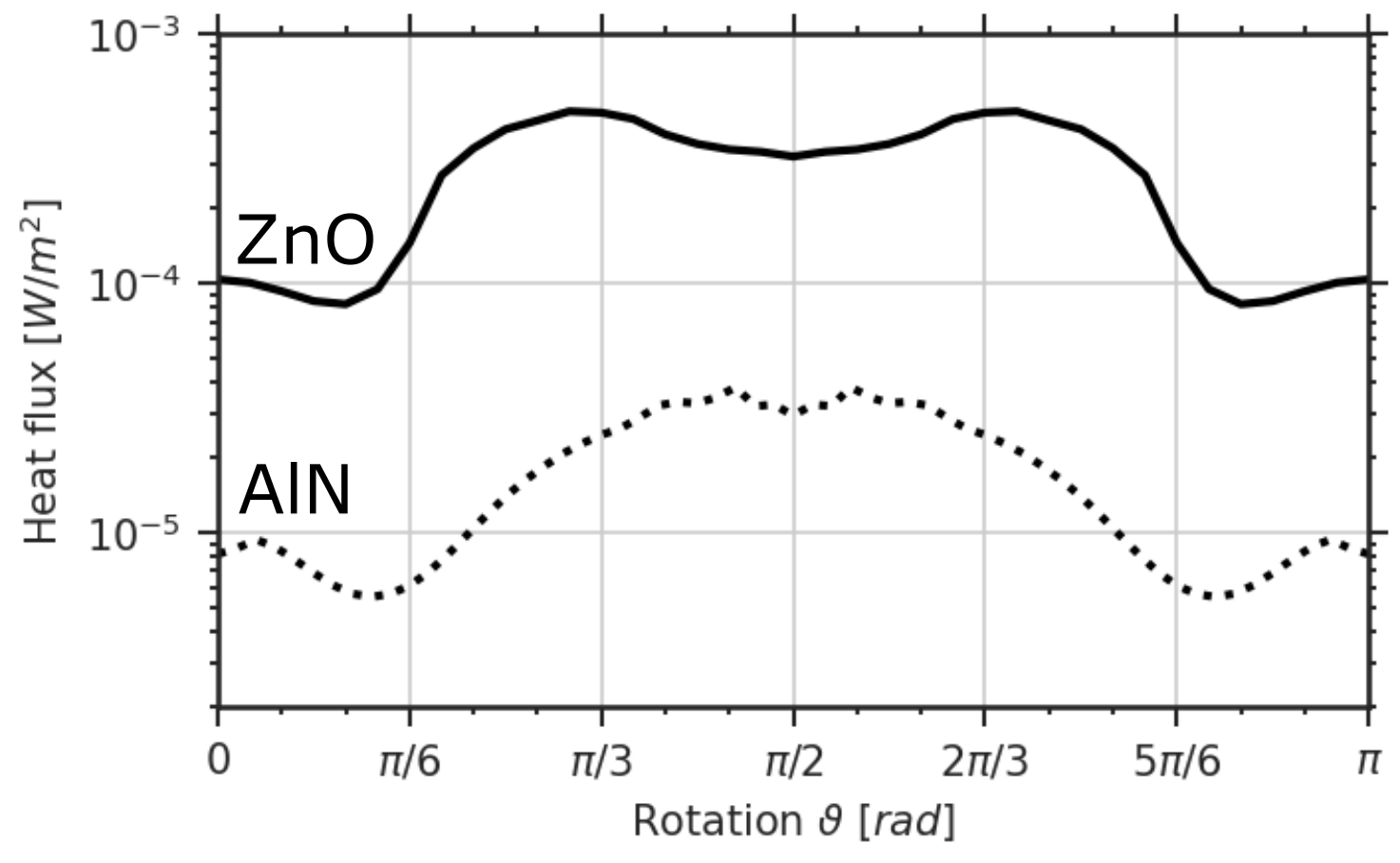}
	\caption{Comparison of emitted PEMHT as a function of the crystal rotation angle $\vartheta$ for ZnO (solid line) and AlN (dotted line). The heat fluxes are calculated for a gap width of $1$ nm at a temperature of $0.1$ K.}
	\label{fig:orientations}
\end{figure}

Next, we plot in Fig.\ref{fig:orientations} the emitted PEMHT for two different piezoelectric materials, ZnO and AlN\cite{Tsubouch1985},  as a function of the orientation angle $\vartheta$, with a fixed gap width of $1$ nm and a temperature of $0.1$ K, to illustrate the influence of the material choice and orientations.

Obvious similarities are shared by the heat fluxes of ZnO (solid line) and AlN (dotted line). For example, for both materials the flux is generally stronger around the orientation $\vartheta=\pi/2$, contrasting to those around $0$ and $\pi$, and sharp slopes appear around $20^\circ$ and $40^\circ$ leading to more than five-fold, step-like variations of the heat flux. These similarities come about because both materials have the $6mm$-wurtzite crystal symmetry. The underlying physics of these features is that as $\vartheta$ is the angle between the normal of the solid-vacuum interface and the crystal's piezoelectric axis (the $c$-axis), and when $\vartheta$ is close to $90^\circ$, the $c$-axis is more aligned with the surface. As a result, the reflected evanescent modes, which propagate only on the surface, can excite a stronger piezoelectric response, leading to an enhanced electrostatic coupling across the vacuum. It has been shown that even complete tunneling of certain acoustic modes can be achieved for ZnO, when $\vartheta$ is between $60^\circ$ and $120^\circ$\cite{Geng2022_2}. 

In addition, PEMHT for ZnO is about one order of magnitude stronger than that of AlN. This is mostly due to the differences in their phase velocities. When the characteristic wavelength $\lambda_T$ is much larger than the gap width, the heat flux scales roughly with $v_\alpha^{-3}$. As a result, a ten-fold difference in the heat flux is expected since AlN has an average phase velocity $v\approx7500$ m/s, whereas it is about $v\approx3900$ m/s for ZnO.

To summarize, we presented a general formulation for piezoelectrically mediated heat transfer (PEMHT) via acoustic phonon tunneling
. Such a formulation can be used to compute the heat flux across a vacuum gap, between arbitrarily anisotropic and oriented piezoelectric crystals. Our analytical and numerical studies reveal that PEMHT provides a significant heat flux, generally more than three orders of magnitude stronger than that from the blackbody radiation when the gap width $d$ is smaller than the phonon characteristic thermal wavelength $\lambda_T$. By comparing to other near-field mechanisms, our study shows that PEMHT can dominate the heat flux  at temperatures below 50 K.

Finally we remark that with the currently available nano-fabrication and cryogenic measurement techniques, PEMHT could possibly be investigated experimentally. We believe that further understanding and engineering of acoustic phonon tunneling can be crucial for many application areas such as nano-electronics, low temperature detectors, quantum information devices and others, for which increasing demand for heat manipulation and management exists.  

This study was supported by the Academy of Finland project number 341823.

	\nocite{*}
	
	\bibliography{references}
\end{document}